\documentclass[copyright,creativecommons]{eptcs}
\usepackage{listings}
\usepackage{xcolor}
\usepackage{graphicx}
\usepackage{wrapfig}
\usepackage{caption}
\usepackage{tabu}
\usepackage{xspace}

\colorlet{light-gray}{gray!20}

\definecolor{dkgreen}{rgb}{0,0.4,0}
\definecolor{gray}{rgb}{0.4,0.4,0.4}
\definecolor{black}{rgb}{0.0,0.0,0.0}
\definecolor{mauve}{rgb}{0.58,0,0.82}
\definecolor{orange}{rgb}{0.7,0.30,0}
\definecolor{dkblue}{rgb}{0.0,0.0,0.4}
\definecolor{ltyellow}{rgb}{1.0,1.0,0.92}
\definecolor{red}{rgb}{0.4,0.0,0.0}
\definecolor{ltred}{rgb}{0.7,0.0,0.0}
\definecolor{pp}{rgb}{0.0,0.4,0.0}
\definecolor{linkblue}{rgb}{0.0,0.0,1.0}

\hypersetup{%
   colorlinks=false,
   urlbordercolor=linkblue,
   pdfborderstyle={/S/U/W 0.5}
}

\newcommand{\Xlibname}[1]{\texttt{#1}}
\newcommand{\Xtopicname}[1]{\texttt{#1}}
\newcommand{\Xbookslist}{\href{http://groups.google.com/forum/\#!forum/acl2-books}{ACL2-Books}\xspace}

\newcommand{\Xacldoc}{ACL2-Doc\xspace}

\newcommand{\Xtopiclink}[2]{\href{http://www.cs.utexas.edu/users/moore/acl2/current/combined-manual/?topic=#1}{\Xtopicname{#2}}}
\newcommand{\Xtextlink}[2]{\href{http://www.cs.utexas.edu/users/moore/acl2/current/combined-manual/?topic=#1}{#2}}
\newcommand{\XCombinedManual}{\href{http://www.cs.utexas.edu/users/moore/acl2/current/combined-manual}{ACL2+Books Manual}\xspace}

\newcommand{\Xdef}[1]{{\ttfamily\color{blue}{#1}}}
\newcommand{\Xfn}[1]{{\ttfamily\color{mauve}{#1}}}
\newcommand{\Xid}[1]{{\ttfamily\color{dkgreen}{#1}}}
\newcommand{\Xkwd}[1]{{\ttfamily\color{orange}{#1}}}
\newcommand{\Xtag}[1]{{\ttfamily\color{ltred}{#1}}}
\newcommand{\Xpp}[1]{{\ttfamily\color{pp}{#1}}}
\newcommand{\Xstr}[1]{{\ttfamily\color{red}{#1}}}
\newcommand{\Xcmt}[1]{{\ttfamily\color{gray}{#1}}}
\newcommand{\Ycmt}[1]{{\rmfamily\color{gray}{#1}}}

\providecommand{\event}{ACL2 2014} 

\title{Industrial-Strength Documentation for ACL2}
\author{Jared Davis
\institute{Formal Verification Group \\
Centaur Technology}
\email{jared@centtech.com}
\and
Matt Kaufmann
\institute{Department of Computer Science \\
University of Texas at Austin}
\email{kaufmann@cs.utexas.edu}
}
\def\titlerunning{Industrial-Strength Documentation for ACL2}
\def\authorrunning{Jared Davis and Matt Kaufmann}
\begin{document}
\maketitle

\begin{abstract}

The ACL2 theorem prover is a complex system.  Its libraries are vast.
Industrial verification efforts may extend this base with hundreds of thousands
of lines of additional modeling tools, specifications, and proof scripts.
High quality documentation is vital for teams that are working together on
projects of this scale.
We have developed \emph{XDOC}, a flexible, scalable documentation tool for ACL2 that can
incorporate the documentation for ACL2 itself, the Community Books, and an
organization's internal formal verification projects, and which has many
features that help to keep the resulting manuals up to date.
Using this tool, we have produced a comprehensive, publicly available ACL2+Books
Manual that brings better documentation to all ACL2 users.  We have also
developed an extended manual for use within Centaur Technology that extends the
public manual to cover Centaur's internal books.  We expect that other
organizations using ACL2 will wish to develop similarly extended manuals.


\end{abstract}

\section{Introduction}

Since its earliest versions, the ACL2 theorem prover has featured a
comprehensive user's manual.  You can find references to this documentation
even as far back as the release notes for ACL2
\Xtextlink{ACL2____NOTE1}{version 1.1}, dating from October 1990.  This
documentation was initially available only from the terminal, but by 1994,
Kaufmann and Moore~\cite{94-kaufmann-design-goals} reported that it had grown
to feature a markup language and conversion tools for producing manuals:
\begin{quotation}
 ``The ACL2 documentation is maintained in a hypertext-like structure which may
  be browsed via ACL2 documentation commands.  In addition, it may be browsed
  via Emacs' Info and via Mosaic.  Roughly .9 megabytes of online documentation
  is available about ACL2, including tutorials.  Instructional materials are
  being prepared as part of the documentation.  The ACL2 user may wish to
  document his or her formal models and browse that documentation with the
  facilities provided...''
\end{quotation}
This mention of Mosaic is amusing, but lends some historical context.
Milestones like the first draft specification of HTML and the first
alpha of Mosaic had each occurred only about a year earlier.  Browsers
like Netscape were not yet available.

This documentation system---now called the \emph{legacy documentation
  system}---went on to serve, largely unchanged, as the basis for the ACL2
User's Manual for the next twenty years.  This time period spanned dozens of
ACL2 releases and thousands of changes to ACL2.  Throughout these changes,
great care was taken by Kaufmann and Moore to expand the manual and keep its
contents up to date; the manual tripled in size from ACL2 Version 1.9 in 1994,
the earliest public release, through Version 6.3 in 2013.

\begin{figure}[p]
\noindent \begin{tabular}{ll}
\includegraphics[width=3in]{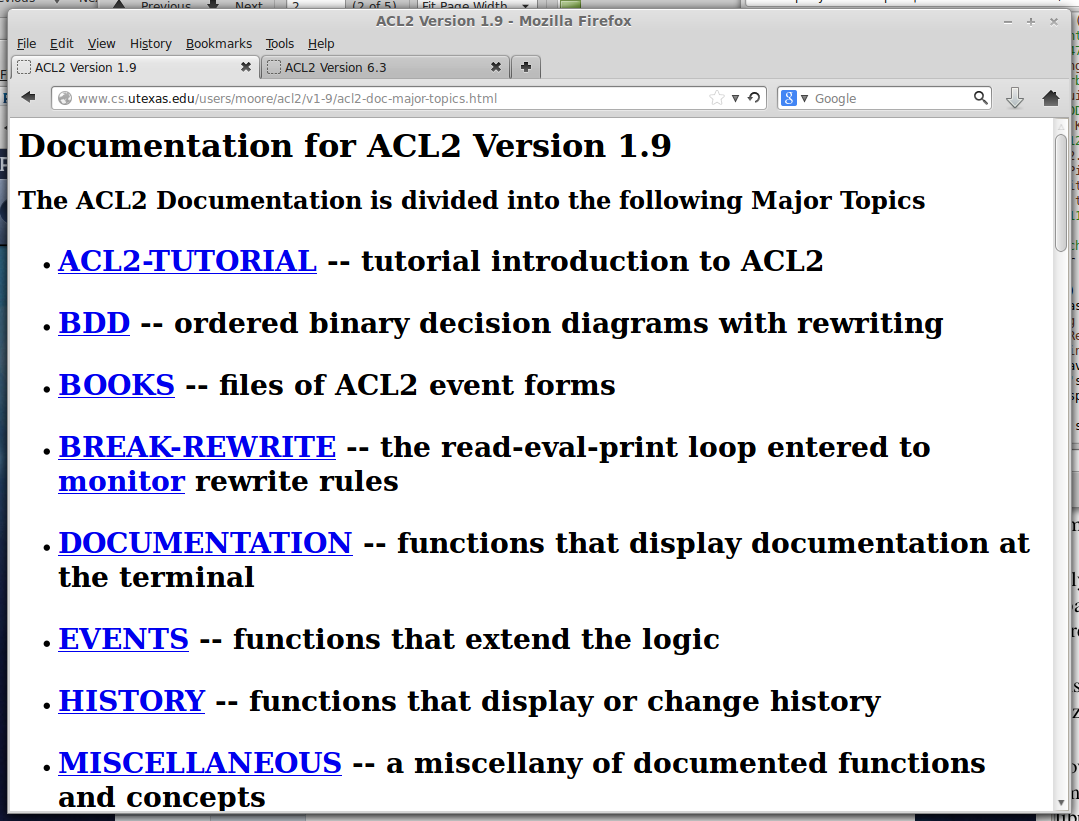}
&
\includegraphics[width=3in]{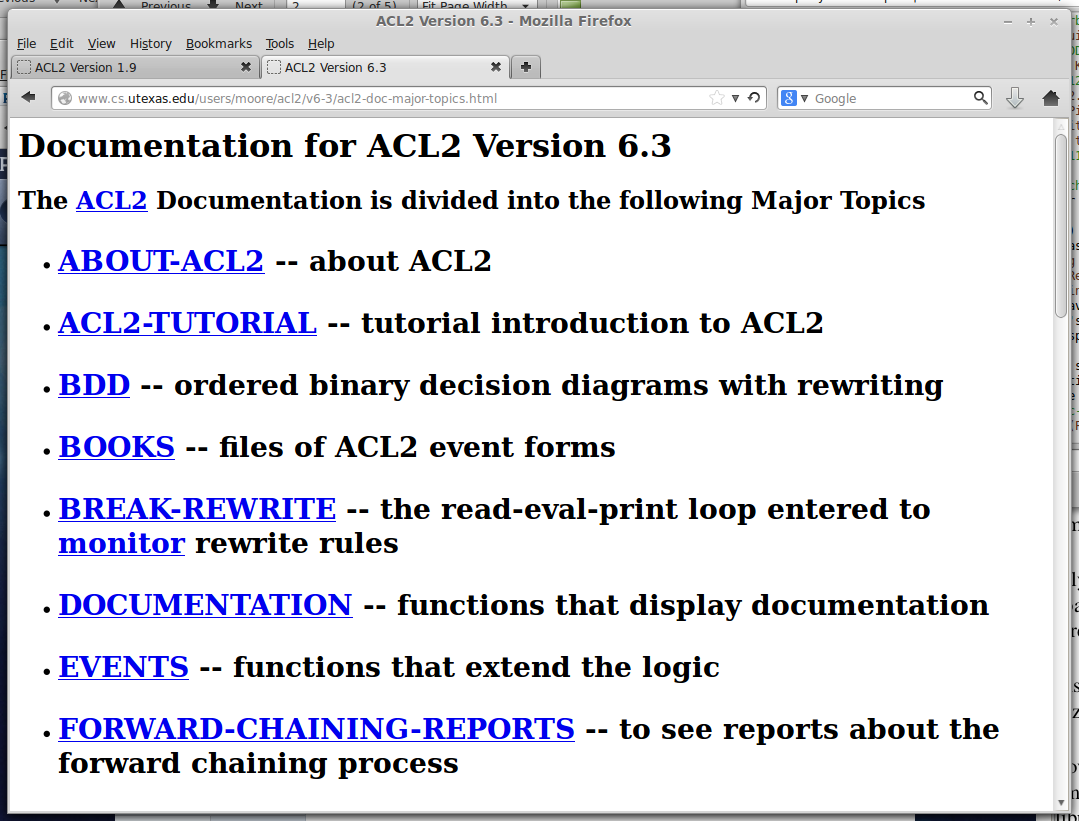} \\
\includegraphics[width=3in]{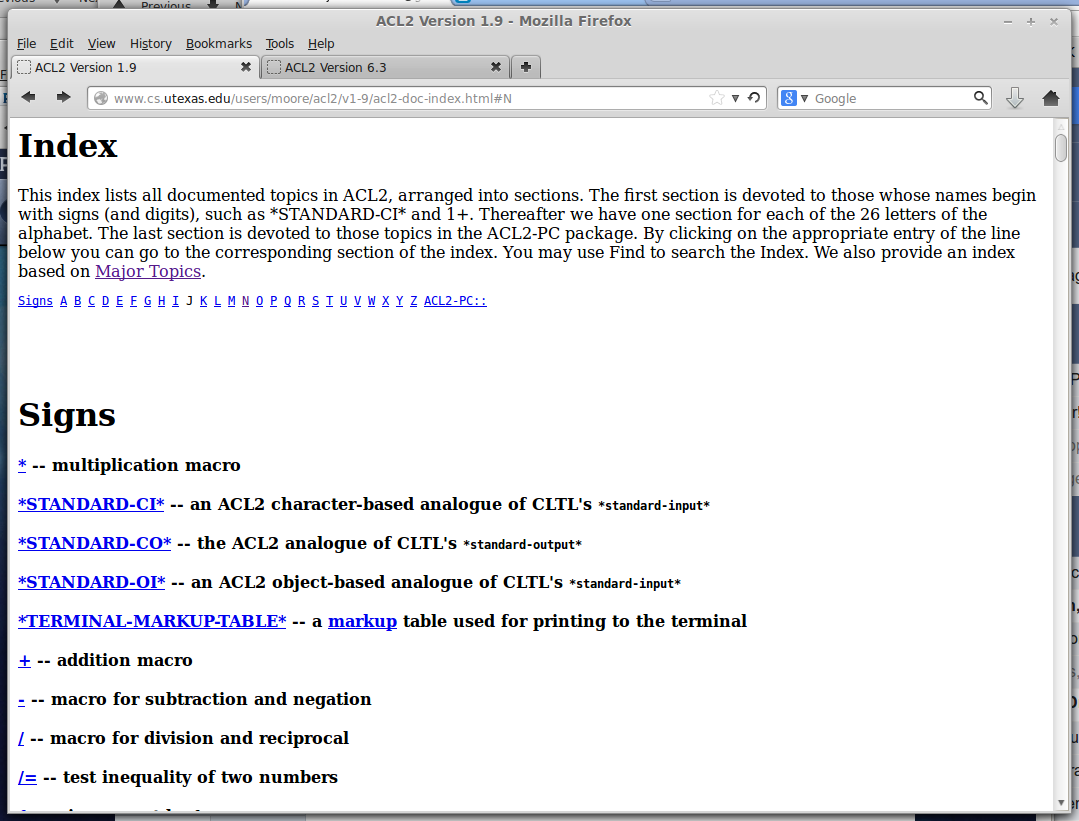}
&
\includegraphics[width=3in]{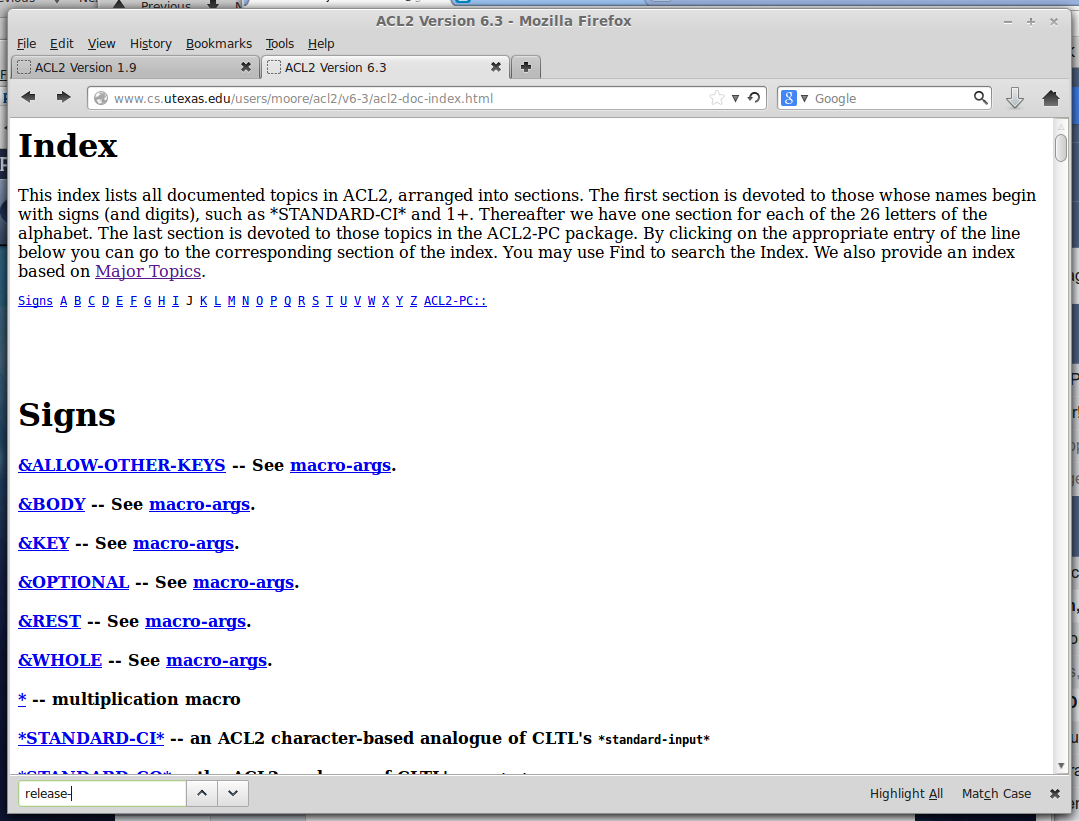}
\end{tabular}
\caption{ACL2 User Manual navigation pages for versions 1.9 (left) and 6.3
  (right).}
\label{fig:legacy-doc}
\end{figure}

\begin{figure}[p]
\noindent \begin{tabular}{ll}
\includegraphics[width=3in]{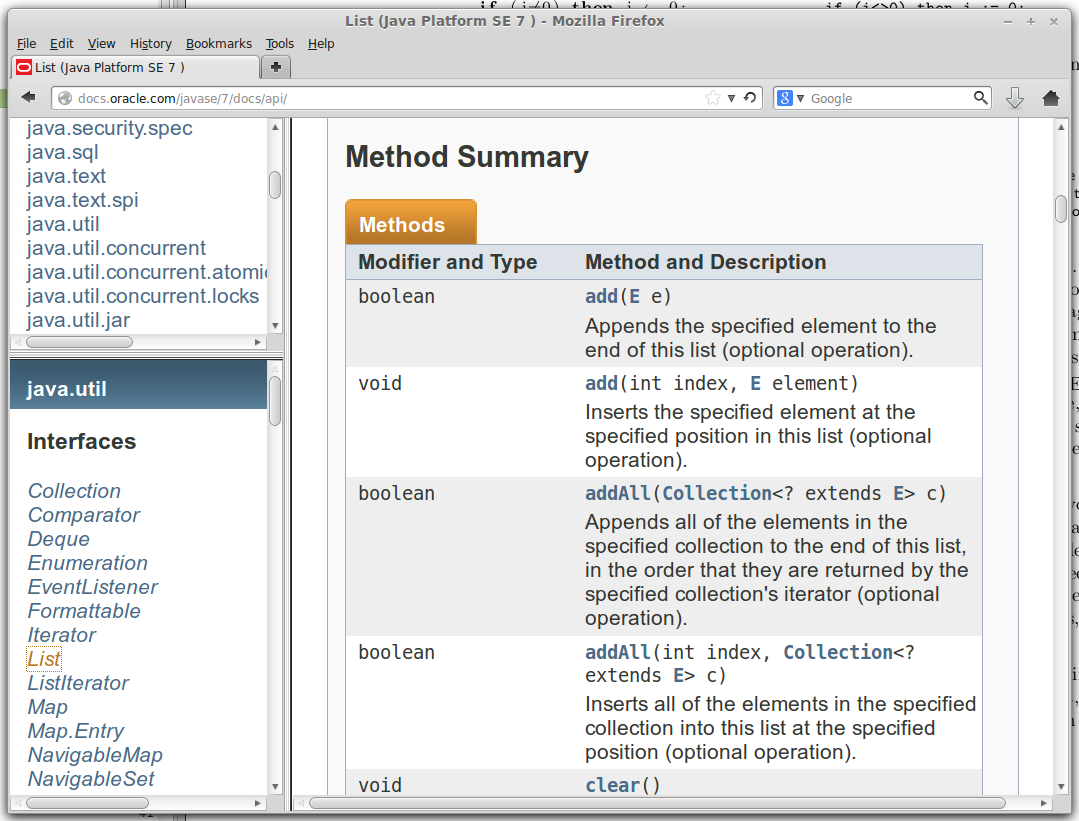}
&
\includegraphics[width=3in]{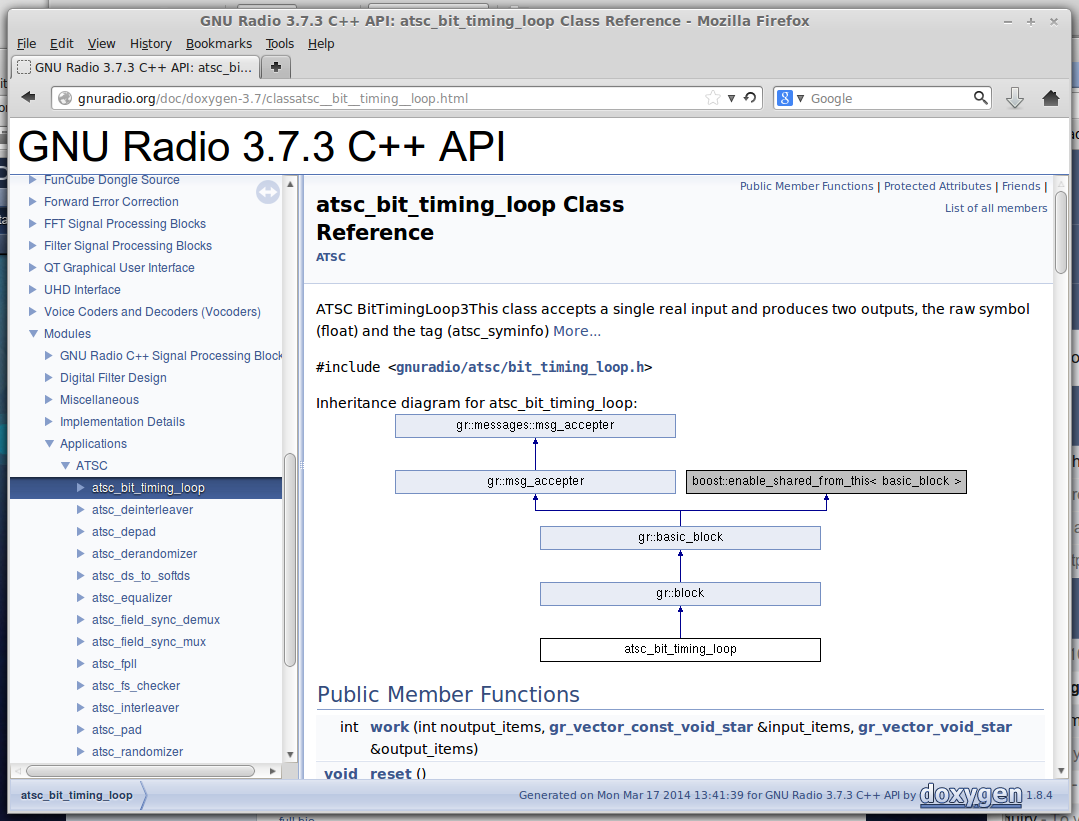}
\end{tabular}
\caption{Java API manual from Javadoc (left) and GNU Radio manual from Doxygen (right).}
\label{fig:java-doc}
\end{figure}

But meanwhile, as web technologies advanced, the presentation, navigation, and
organization of the manual were not the focus.  In Figure \ref{fig:legacy-doc},
you can see the great similarity between the key navigation pages from Versions
1.9 and 6.3.  By way of contrast, Figure \ref{fig:java-doc} shows manuals
produced by the
\href{http://www.oracle.com/technetwork/java/javase/documentation/javadoc-137458.html}{Javadoc}
and \href{http://www.doxygen.org/}{Doxygen} tools, which, for instance, take
advantage of frame-based navigation.

Moreover, throughout this time, the manual only covered the functionality
provided by the theorem prover itself.  While many topics mentioned related
\href{http://acl2-books.googlecode.com/}{Community Books} and
\href{http://www.cs.utexas.edu/users/moore/acl2/workshops.html}{workshop
  papers}, there were no direct links.  Even well-established libraries like
\Xlibname{data-structures} and \Xlibname{ihs}, which had documentation, were
excluded from the manual.  The tools for creating manuals for books were
difficult to use and had various problems with packages.  Jared Davis described
these and other frustrations with the legacy documentation system in a
September 3, 2009 email to Matt Kaufmann and Alan Dunn.  He explained his
motivation as follows:
\begin{quotation}
 ``At Centaur we are developing a lot of books, and we are making use of lots of
  books in \texttt{:dir :system}.  We have a number of users, with varying
  levels of ACL2 proficiency, and varying levels of understanding of each
  library.  In some cases they may not even know what libraries they are using.
  Because of this, it would be really nice to be able to point them to
  documentation for our whole system, instead of just pieces of it...  [This]
  does not work.''
\end{quotation}
Dunn was able to quickly patch the legacy documentation system to correct these
package problems, allowing Centaur to use it, in the short term, in their X86 processor
verification efforts.  But to address
the deeper issues, we began working on a new documentation system for ACL2
named XDOC.  On October 20, we committed a
\href{https://code.google.com/p/acl2-books/source/detail?r=383}{preliminary
  version} to the Community Books repository.  In the years since, we have
grown XDOC into a mature and capable documentation tool for ACL2.

This paper is a broad introduction to XDOC for the ACL2 community.  It provides
some historical context and describes our motivation for developing a new
documentation system.  It explains how XDOC impacts the ACL2 User's Manual, and
the new options for viewing ACL2's documentation.  It shows how you can use
XDOC to document your own books, and how to do this in elegant ways that can
reduce the burden of writing and maintaining documentation.  It also describes
how organizations can effectively use XDOC to develop comprehensive manuals for
their verification projects.

\begin{figure}[b!]
\noindent \begin{tabular}{ll}
\includegraphics[width=3in]{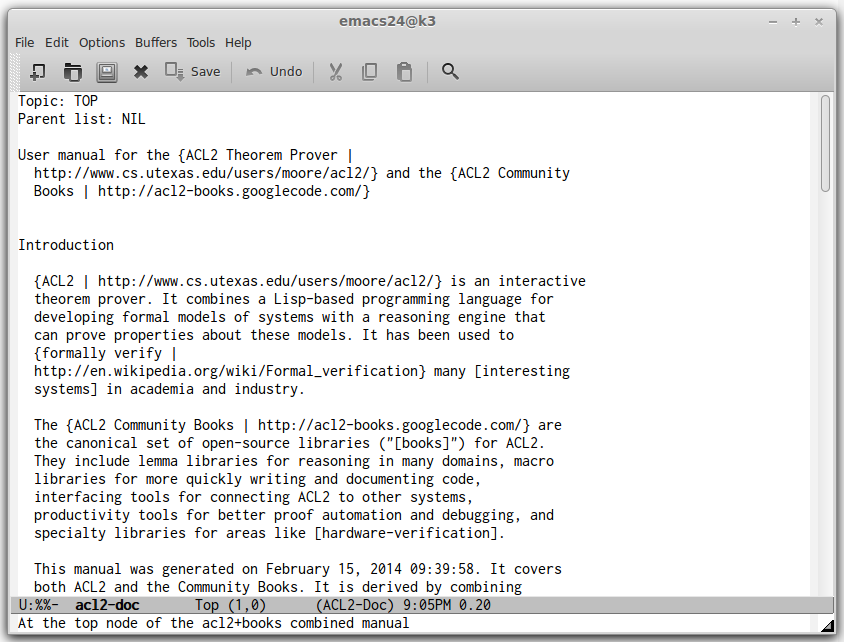}
&
\includegraphics[width=3in]{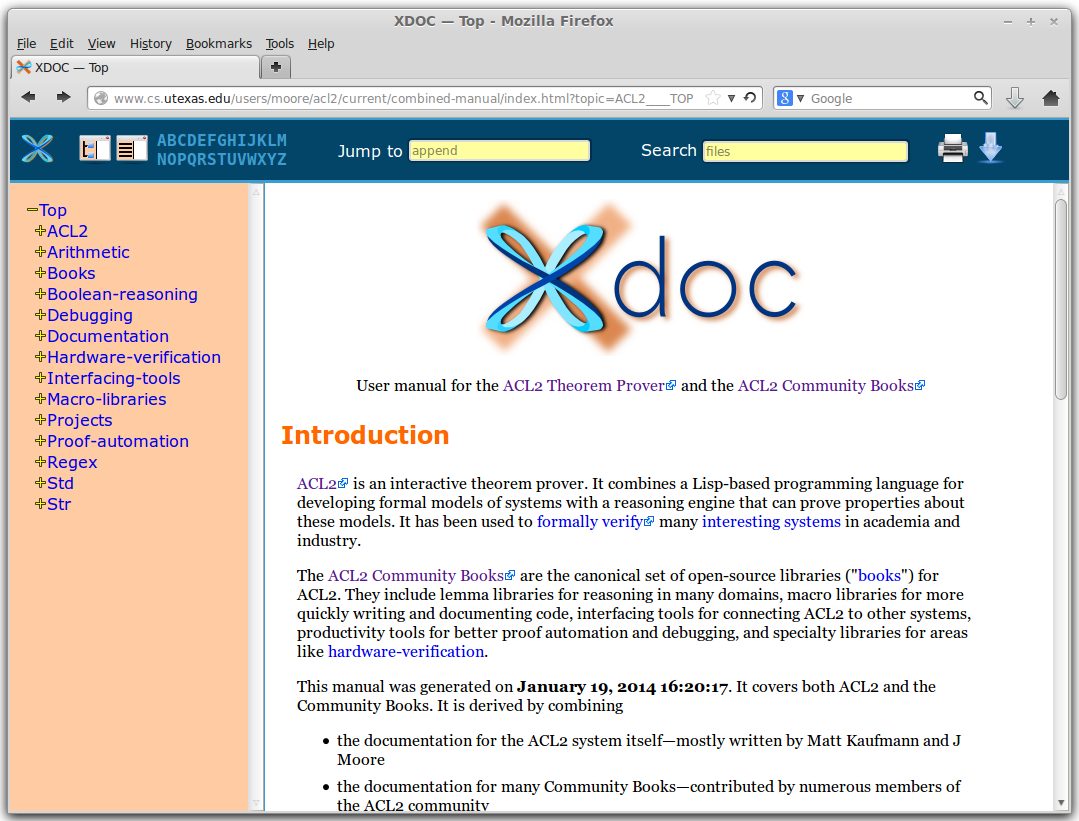}
\end{tabular}
\caption{XDOC manuals in the Emacs Viewer (left) and the Fancy Viewer (right).}
\label{fig:viewers}
\end{figure}

Just what is XDOC?  Like many software documentation tools, it allows pieces of
documentation---describing a function, a theorem, or what have you---to be
scattered throughout the source code of many books.  This documentation can
make use of an XML-based \Xtextlink{XDOC____MARKUP}{markup language} and
also a \Xtextlink{XDOC____PREPROCESSOR}{\emph{preprocessor}} that has a
number of features.  We describe the basics of writing XDOC documentation in
Section \ref{sec:usage}.


XDOC can assemble these disparate topics to create a manual for a fancy
web-based viewer or for an Emacs-based viewer, both shown in Figure
\ref{fig:viewers}.  These manuals can be very comprehensive.  Starting with
ACL2 version 6.4, we have converted the documentation for ACL2 itself from its
legacy format into XDOC format.  Meanwhile, over the past few years, many
Community Books have been extended with XDOC documentation.  In Section
\ref{sec:combined}, we explain how these efforts have allowed us to create a
single, integrated \XCombinedManual that covers both ACL2 itself and a large
collection of Community Books.  We also describe the options for accessing this
manual, and explain how it can be extended to cover additional books.

For industrial verification projects, XDOC can also create extended manuals
that cover additional books that are internal to an organization.  For
instance, the formal verification group at Centaur Technology now has a custom
manual that extends the ACL2+Books Manual with thousands of additional topics
that are specific to their particular modeling and verification efforts.  This
manual is built nightly and is available company-wide as an intranet web page.
David Rager has told us that a similar manual is being used internally at
Oracle.  We explain how to create extended manuals in Section \ref{sec:custom}.

A major reason to embed the documentation in the source code is that this makes
it easier to keep the manual updated as the source code is changed.  Good
software documentation tools can also automatically extract and use information
from the source code, e.g., function signatures, class hierarchies, etc., to
provide context, generate navigation pages, and so on, which avoids duplication
between the manual and the code and helps to keep the manual current.
Unfortunately, features of ACL2 such as dynamic typing and macros, and also its
lack of any built-in notion of module, conspire to make this kind of automatic
extraction difficult.  In Section \ref{sec:automation} we explore this
difficulty and describe some solutions.  We have developed some basic tools for
automatically grouping related events and including them in a documentation
topic.  We have also been able to develop documentation-aware macros that can
produce high quality documentation without duplicating code.  To highlight
this, we show how certain macros in the
\Xtopiclink{ACL2____STD_F2UTIL}{\Xlibname{std/util}} library are now integrated
with XDOC.

\section{Writing Documentation with XDOC}
\label{sec:usage}

As groundwork for reading the rest of this paper, it will be very
useful to have a basic feel for how XDOC documentation is written.  In
this section, we dissect an example topic that illustrates the basics.

Since XDOC is a library, even its most basic documentation commands are not
included in ACL2 itself.  The first step to using XDOC is to load it via:
\begin{lstlisting}[language=Lisp]
(^\Xdef{include-book}^ "xdoc/top" ^\Xkwd{:dir :system}^)
\end{lstlisting}


You may find that you already have this book loaded, as it is now widely
included throughout the Community Books.  If not, it loads very quickly, does
not alter your theory, and is \Xtextlink{ACL2____DEFTTAG}{trust-tag} free.  Once it is loaded, we can begin
writing documentation.  The basic unit of documentation in XDOC is a
\emph{topic}, and the most primitive way to add a new topic is called
\Xtopiclink{ACL2____DEFXDOC}{\Xdef{defxdoc}}.  Here is an example:
\begin{lstlisting}[language=lisp]
(^\Xdef{defxdoc}^ ^\Xfn{getopt}^
  ^\Xkwd{:parents}^ (interfacing-tools)
  ^\Xkwd{:short}^ "A library for processing command-line options."
  ^\Xkwd{:long}^ "^\Xtag{<h3>}^Introduction^\Xtag{</h3>}^

^\Xtag{<p><b>}^Getopt^\Xtag{</b>}^ is a tool for writing command-line programs in ACL2.  It is
similar in spirit to
^\Xtag{<a href='http://perldoc.perl.org/Getopt/Long.html'>}^Getopt::Long^{\Xtag{</a>}^ for Perl,
^\Xtag{<a href='http://trollop.rubyforge.org/'>}^Trollop^\Xtag{</a>}^ for Ruby, and so on.^\Xtag{</p>}^

^\Xtag{<p>}^We basically extend ^\Xpp{@(see defaggregate)}^ with a command-line parsing layer.
This has some nice consequences:^\Xtag{</p>}^ ...")
\end{lstlisting}

\noindent As a quick explanation:

\begin{itemize}

\item Each topic has a unique name which must be an ACL2 symbol---in this case
  \Xfn{getopt}.  Since ACL2 symbols belong to particular packages, the usual
  package mechanism can be used to avoid name clashes when combining
  documentation from many libraries.

\item The \Xkwd{parents} refer to other topics that should typically
  cover broader or more general concepts.  This parent information is
  used to group documentation into a hierarchy that supports basic
  navigation features in viewer programs.  Most topics have only a
  single parent, but having additional parents can be useful when a
  topic fits well in many places.  For instance, it allows the
  \Xtopiclink{ACL2____STD_F2UTIL}{\Xlibname{std/util}} library to be
  listed as part of the \Xtopiclink{ACL2____STD}{\Xlibname{std}}
  library, and also under the
  \Xtopiclink{ACL2____MACRO-LIBRARIES}{macro-libraries} category.

\item The \Xkwd{short} and \Xkwd{long} strings, together, make up the real
  contents of a topic.  The short string is intended to be a 1-2 sentence
  summary of the topic.  Viewers embed it into navigational devices such as
  tooltips, child topic listings, search results, and so on.  The long string
  contains the main contents of the topic.  Both strings are optional, which
  is useful for topics that need little explanation.


\end{itemize}

The short and long strings can make use of a conventional, XML markup language
with tags like \Xtag{<p>}, \Xtag{<h3>}, etc.  This language is very similar to
a simple subset of HTML except that the tags must be strictly balanced.  XML is
very widely supported.  For instance, web browsers can directly load XDOC's
markup and, e.g., traverse it using Javascript, or convert it into HTML using
XSLT stylesheets.  More generally, almost every programming language has libraries
for parsing and working with XML files.




In addition to XML tags, the short and long strings can also make use of
\emph{preprocessor directives} such as \Xpp{@(see defaggregate)}.  The
preprocessor is a central feature of XDOC.  It is used to translate these
special \Xpp{@} directives into plain XML when you create a manual or display
an individual topic with the \Xkwd{:doc} command.  For instance, in this case
it will convert \Xpp{@(see defaggregate)} into \Xtag{<see
  topic='STD\_\_\_\_DEFAGGREGATE'>\Xstr{defaggregate}</see>}.




The preprocessor can do many things.  For instance, its \Xpp{@(see ...)}
directive handles the details of symbol name encoding and allows topics to be
referenced simply, usually without package prefixes.  Its special
\Xpp{@('\dots')} and \Xpp{@(\{\dots\})} syntax can be used to write raw or
verbatim text without escaping XML characters like \Xtag{<} as \Xstr{\&lt;},
and automatically links to documented topics.  For instance, the translation
of \Xpp{@('(car (append x y))')} will have links to the topics for \Xtopiclink{COMMON-LISP____CAR}{car}
and \Xtopiclink{COMMON-LISP____APPEND}{append}.  Other directives can be useful for automatically
generating documentation and keeping the documentation up to date.  For
instance, its \Xpp{@(def ...)}  directive can automatically look up functions,
theorems, etc., from the ACL2 world, and insert them into the documentation as
code blocks with automatic links to any documented topics.  Other directives
allow for evaluating ACL2 expressions and injecting the results into the
documentation, which may be useful for ensuring that documented examples are
correct.

The \Xdef{defxdoc} command is a real ACL2 \Xtextlink{ACL2____EVENTS}{event}.
Many other documentation tools, such as Javadoc and Doxygen, instead use
special comments like \Xcmt{/**~...~*/} to embed the documentation into the
source code.  These comments can be extracted by the documentation tool and
associated with nearby source code elements, but are invisible to the Java or
C++ compiler.  This means it isn't possible to dynamically manipulate existing
documentation.

In contrast, XDOC stores all of its topics in an ordinary ACL2 \Xtextlink{ACL2____TABLE}{table}, and
\Xdef{defxdoc} just adds a new topic to this table.  One consequence of this is
that, when interacting with ACL2, all of the documentation for the books you
have loaded and the events you have submitted is immediately available using
the usual \Xkwd{:doc} command.  For instance, after submitting the
\Xfn{getopt} topic above, we can display it as text:


\begin{lstlisting}[language=Lisp]
ACL2 !> (^\Xdef{defxdoc}^ ^\Xfn{getopt}^ ...)
ACL2 !> ^\Xkwd{:doc}^ getopt
ACL2::GETOPT -- Current Interactive Session
Parents: INTERFACING-TOOLS.

  A library for processing command-line options.


Introduction

  Getopt is a tool for writing command-line programs in ACL2. It is
  similar in spirit to {Getopt::Long |
  http://perldoc.perl.org/Getopt/Long.html} for Perl, {Trollop |
  http://trollop.rubyforge.org/} for Ruby, ^\texttt{and}^ so on.

  We basically extend [defaggregate] with a command-line parsing layer.
  This has ^\texttt{some}^ nice consequences:
  ...
\end{lstlisting}

Having the documentation in a table also means that we can easily write code to
extend existing topics, change parent relationships, and so forth.  In Section
\ref{sec:automation}, we will see some interesting and useful ways to take
advantage of these capabilities.


\section{The ACL2+Books Manual}
\label{sec:combined}

Starting in ACL2 6.4, the documentation for ACL2 has been converted into XDOC
format, made editable by the Community, and integrated with the documentation
for the Community Books to create a convenient ACL2+Books Manual.  In this
section, we describe some of the history leading up to this change, the new
options available for viewing this manual, and how members of the ACL2
Community can add new documentation to the manual.

\subsection{Steps Toward a Unified Manual}

Although the initial version of XDOC (October 2009) lacked many basic features
such as navigation pages, its XML markup language and preprocessor were similar
to the XDOC of today.  Within a few months, we added a basic navigation system
and the formal verification group at Centaur began using XDOC.  Over the next
year, many libraries were documented with it.

Meanwhile, ACL2 itself and many other libraries were still documented with the
legacy system.  Frustrated with this schism, Warren Hunt challenged us to unify
the manuals.  Toward this goal, we developed a tool for incorporating legacy
documentation into XDOC and implemented an \Xkwd{:xdoc} command for
interactive browsing.  In January 2011, we published an extremely preliminary
ACL2+Books Manual and
\href{https://groups.google.com/d/topic/acl2-books/KfSYZUf08ns/discussion}{announced}
it to the \Xbookslist email list.  Over the next years, as ever more XDOC
documentation became available, the convenience of a combined manual became
increasingly apparent.  With the release of ACL2 5.0 in August 2012, Centaur
began hosting \href{http://fv.centtech.com/acl2/}{unofficial ACL2+Books
  Manuals}.


The tipping point came in July 2013 when we developed a new, fancy web-based
XDOC viewer.  The new viewer included many expanded navigation options, e.g.,
jumping to topics by name, flat and hierarchical navigation, and the ability to
expand subtopics inline within a page.  With such an improved viewer, it seemed
that the time to retire the legacy system had come.  In September, Davis
\href{http://www.cs.utexas.edu/users/moore/acl2/seminar/2013.09.20-jared/xdoc.pdf}{presented XDOC} to the \href{http://www.cs.utexas.edu/users/moore/acl2/seminar/index.html}{ACL2 Seminar} at UT, and proposed converting ACL2's
documentation into XDOC format and making it editable by the community to
better integrate the system- and book-level documentation.

After developing some custom tools to automate the process, we carried out this
conversion and developed a way to load the resulting documentation---now
separate from ACL2---back into the theorem prover at build time, so that the
\Xkwd{:doc} command would continue to function even without including
any of the XDOC books (which convert \Xkwd{:doc} to \Xkwd{:xdoc} using
the \Xtopiclink{ACL2____LD-KEYWORD-ALIASES}{ld-keyword-aliases}
feature of ACL2).  One last sticking point was that XDOC did not
support browsing the documentation in Emacs.  We developed a new Emacs
browser, named \Xacldoc, that is easy to use, provides several useful
capabilities not present in its predecessor that was based on Emacs
Info, and can automatically download a gzipped copy of the ACL2+Books
manual.

With the technical hurdles out of the way, on October 27, 2013 we
announced the new manual to the general ACL2 mailing list.  Since
then, we have begun to take advantage of XDOC features in ACL2
documentation topics, and to more closely integrate the
system and book documentation.  For instance, we have extended topics
for many simple functions like
\Xtopiclink{COMMON-LISP____APPEND}{append} with \Xpp{@(def ...)}
directives so that the documentation shows their definitions.  In many
cases we have added new links from ACL2 topics into related library
documentation.  We have also worked to improve the organization of
topics, e.g., new sections like
\Xtopiclink{ACL2____INTERFACING-TOOLS}{interfacing-tools} allow
system-level topics like \Xtopiclink{ACL2____IO}{io},
\Xtopiclink{ACL2____SAVE-EXEC}{save-exec} and
\Xtopiclink{ACL2____SYS-CALL}{sys-call} to be grouped with libraries
such as \Xtopiclink{ACL2____GETOPT}{getopt},
\Xtopiclink{ACL2____OSLIB}{oslib}, and
\Xtopiclink{ACL2____QUICKLISP}{quicklisp}.  We have also worked to
move many topics out of patchwork areas like
\Xtopiclink{ACL2____MISCELLANEOUS}{miscellaneous} and
\Xtopiclink{ACL2____OTHER}{other} and into more germane sections.


\subsection{Viewing the Manual}

There are currently three main ways to view documentation: the interactive
\Xkwd{:doc} command, the fancy web-based viewer, and the \Xacldoc viewer for
Emacs.

The \Xkwd{:doc} command works much as it always has.  Although it is a
text-only display that lacks, e.g., colors, clickable links, and so forth, we
often find it useful for quick reference.  The most important thing to note is
that it will \textbf{only} display documentation for (1) the theorem prover itself,
and (2) books that you have loaded.  This has always been true, even for the
legacy documentation system, but the effect is now more pronounced as so many
books now have documentation.

We expect that most users will want to use the fancy
\href{http://www.cs.utexas.edu/users/moore/acl2/current/combined-manual}{web-based
  edition} of the ACL2+Books Manual, which is available on the
\href{http://www.cs.utexas.edu/users/moore/acl2/}{ACL2 web site}.  We plan to
update this manual with each ACL2 release, as has historically been done
for the ACL2 User's Manual.  For users of bleeding-edge, development versions
of ACL2 and the Community Books, there is also a frequently updated
\href{http://www.cs.utexas.edu/users/moore/acl2/manuals/current/manual}{documentation
  snapshot}.

\begin{wrapfigure}[6]{r}[0pt]{2.5in}
\vspace{-0.5em}
\fbox{\includegraphics[width=2.3in]{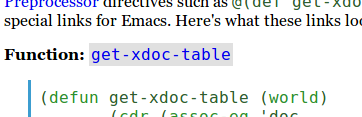}}
\end{wrapfigure}

While using the web-based version should be largely self-explanatory, a couple of
features are worth noting.  If you use Emacs, you may be able to configure your
browser so that clicking on special gray \Xtextlink{XDOC____EMACS-LINKS}{\emph{Emacs links}} (as shown to the
right) will automatically trigger a tags search---this can be very
convenient, but requires some
setup; for details, see the documentation on XDOC.  If you have a slow or
intermittent network connection, every XDOC manual includes a \emph{Download
  this Manual} button which makes it easy to get a local copy for offline
browsing.  Finally, the \includegraphics[width=2.5mm]{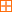} button next
to \emph{Subtopics} listings can be used to expand whole sections of the manual
into a single page, which can be especially useful when reading or printing tutorial-style
material.

The third way to view documentation is with the
\Xtextlink{ACL2____ACL2-DOC}{\Xacldoc} Emacs browser.  This view is
similar to the Emacs Info version of the ACL2 User's Manual in the
legacy documentation system.  However, it has additional capabilities
and, most importantly, it can view not only the ACL2 User's Manual but
also the ACL2+Books Manual.  The text-based view provided by \Xacldoc
is not ideal for objects like tables, icons, and external links.
Emacs users, however, may appreciate the ability to navigate
documentation quickly using familiar Emacs movement commands, without
leaving Emacs or touching the mouse.



\newcommand{\Xemacscmd}[1]{`\texttt{\color{dkgreen}{#1}}'\xspace}

The documentation for \Xacldoc provides a thorough description, so here we give
only a brief introduction.  \Xacldoc is normally loaded automatically by the
usual \texttt{emacs-acl2.el} file, or can be loaded directly from
\texttt{emacs/acl2-doc.el}.  Once loaded, you can invoke it with \Xemacscmd{M-x
  acl2-doc} or \Xemacscmd{C-t g}, or by invoking \Xemacscmd{C-t .} to go to a
particular topic.  The \Xemacscmd{h} command takes you directly to the help
topic, \Xacldoc, which includes a summary of commands.  Single-keystroke
commands allow you to follow a link, to move to the next link on the page, to
search the manual for a string or a regular expression, to search the index, to
see a history of visited topics (which can serve as links), and so on, as well
as to download a bleeding-edge copy of the manual from the web.  Links are
indicated with square brackets, for example, \texttt{[rewrite]}, but typing a
\Xemacscmd{<RETURN>} will take you to the topic at the point even if the brackets
are missing.  The interface is natural for Emacs users, e.g., the \emph{go to
  topic} command offers a default and also supports completion.

\subsection{Contributing to the Manual}

The full contents of the ACL2+Books Manual are now part of the
Community Books.  Everyone in the community is welcome to make
improvements to existing documentation and to extend the manual to
cover additional libraries.  Changes can be discussed and committed at
the
\href{http://acl2-books.googlecode.com}{acl2-books}
project.

The top-level file for the ACL2+Books Manual is found in
\texttt{doc/top.lisp}, and it consists largely of
\Xtopiclink{ACL2____INCLUDE-BOOK}{include-book} commands.  If you
contribute new libraries and would like to have their documentation
included in the manual, it should usually suffice to include your
library into this file.  Similarly, if you add documentation to some
existing Community Books, you may need to ensure they are included.
The documentation for XDOC explains how to build the manual after
making your changes.






If you want to clarify or extend an existing topic, it should usually be easy
to find the right file.  Each of the viewers shows, near the top, the path to
the file where the documentation originates.  Topics that list ``ACL2 Sources''
as their origin can be found in \texttt{system/doc/acl2-doc.lisp}, which
initially consisted of the documentation that we extracted from a copy of the ACL2
sources in October, 2013.

\section{Custom Manuals}
\label{sec:custom}

The ACL2+Books Manual is the most comprehensive publicly available XDOC manual.
However, generally speaking, XDOC can easily create a custom manual for any
particular set of books.

For organizations that are developing large formal verification projects, this
can be especially valuable.  While an industrial verification effort might make
use of many Community Books, it will also typically include many internally
developed books that have not been publicly released, e.g., because they
contain proprietary information or are simply not of general interest.  As the
number of internal books grows, so too does the need for effective
documentation.

To document such a project with XDOC, the typical approach is to create a new
book, say \texttt{doc.lisp}, that simply includes all of the other books in the
project and then invokes the \Xtopiclink{XDOC____SAVE}{xdoc::save} command.
Certifying this book will then produce a new web-based manual.  Such
an internal manual can include all of the documentation for ACL2, the
Community Books you are using, and your internal books.  Deploying a
web-based manual as part of your automatic build process can make it
easy for your team (and organization) to have current documentation
covering your entire project.

A manual produced by \texttt{xdoc::save} is a new directory that
contains a top-level \texttt{index.html} file and other supporting
files.  These manuals do not require any server-side support, so to
deploy one you may only need to upload the directory to your intranet
web server or copy it into some shared file system.  This works well
if all of your users have fast, local intranet connections.  However,
if your organization includes remote users, the basic manual produced
by \texttt{save} may be slow to load: manuals can be tens of
megabytes, and by default all of this content is loaded at startup.  To
avoid this latency, you may wish to deploy a
\Xtextlink{XDOC____DEPLOYING-MANUALS}{server-supported manual}.  XDOC
includes scripts for converting the manual's contents into a database,
and for having a web server access this database to provide topics on
an as-needed basis.  While this is more work, it allows for manuals
that are much more responsive on slow connections.






\section{Automating Documentation and Keeping DRY}
\label{sec:automation}

XDOC, like many software documentation tools, takes the approach of embedding
the documentation directly into the source code.  There are many other ways to
document software.  For instance, documentation could be kept in a Wiki or put
into files such as Microsoft Word documents, \LaTeX{} files, HTML files, etc.
Documentation could even take the form of traditional, paper books, conference
papers, blog articles, presentations, webcasts, and so forth.

These alternatives do have some benefits.  Embedding the documentation in the
source code limits who can practically edit and contribute to it and, at least
to some degree, dictates its form.  For instance, good technical writers and
graphics artists may have little experience with version control systems and
markup languages, and also should perhaps not be directly editing and
committing source code.

Even so, there's a lot to be said for putting the documentation right into the
code.  As a minor benefit, this approach lets developers see the corresponding
documentation, albeit in its raw markup form, whenever they are working with
the code.  It also makes it convenient to update the documentation as the code
is changed: you don't also have to find and edit some corresponding Wiki entry
or file.

But by far the most important benefit of embedding the documentation into the
code is that it can drastically reduce the amount of information that is
duplicated between the code and the documentation.  This is in keeping with
the \emph{don't repeat yourself} (DRY) principle of software development,
popularized by Hunt and Thomas~\cite{99-hunt-pragmatic}.  In practice, we believe
this is critical to keeping the documentation up to date as the
code base changes.  A good documentation tool can avoid this duplication by
automatically extracting information from the source code---function
signatures, type definitions, how code is organized into modules, and so
forth---and making this information available in the manual.

\subsection{Challenges: Modules, Types, and Macros}

In languages like Java and C++, programs are made up of packages, namespaces,
classes, etc., that bundle together related functions and structures.
Documentation tools for these languages can exploit these scoping constructs to
partition the manual into coherent sections automatically, and to automate the
generation of sensible navigation pages.  They can also make good use of the
type declarations in function signatures.  Consider some C++ function from
\href{http://llvm.org}{LLVM}:

\begin{lstlisting}[language=C++]
^\Xid{Pass}^* ^\Xfn{BasicBlockPass}^::^\Xfn{createPrinterPass}^
(^\Xid{raw\_ostream}^ &O, const ^\Xid{std::string}^ &Banner) const
{...}
\end{lstlisting}
Even if the programmer doesn't write anything to explain this function, a
documentation tool still has a lot to work with.  It can link to classes like
\Xid{Pass}, \Xid{raw\_ostream}, and \Xid{std::string}.  It knows that this
function should be filed under the \Xfn{BasicBlockPass} class.  It can link to
the parent classes of \Xfn{BasicBlockPass} and to the \Xfn{createPrinterPass}
method in from the superclass that is being overridden.  This context may be
very useful to a reader who wants to understand what the function does and how
it fits into the program.

It is much harder to provide this kind of automatic context and boilerplate in
a documentation tool for a dynamically typed language like ACL2.  Consider some
Lisp function from ACL2's rewriter:
\begin{lstlisting}[language=Lisp]
(^\Xdef{defun}^ ^\Xfn{geneqv-refinementp}^ (equiv geneqv wrld) ^^
   ...)
\end{lstlisting}
What does this function return?  What kind of object is \texttt{equiv} supposed
to be?  What is a \texttt{geneqv}?  A \texttt{wrld}?  You can discover the
answers by studying the source code comments and understanding the context, but
there is no function signature for a documentation tool to extract.  Worse,
this function is merely one among many.  It's not a member of some class.  It
has no identifiable relationship to any sort of interface.  There is little that
a documentation tool can do to provide useful context, here.

Things may not be quite as bad for user-level ACL2 functions.  Many of these
functions have guards, which are often like type signatures except that they
don't explain the return value.  Meanwhile, ACL2 books may sometimes contain
``type-like'' theorems about the return values of functions.  Perhaps an ACL2
documentation tool could heuristically identify these theorems.  Perhaps it
could also use some kind of call-graph analysis or other kinds of hints---file
names, packages, common name prefixes, and so on---to group up functionality
into related modules.  But this does not seem especially promising.


But there's a deeper challenge here.  Java and C++ programs have a rich but
fixed set of syntactic constructs, like packages, classes, functions,
enumerations, and structs.  In contrast, as ACL2 users, we often introduce our
own custom \Xtopiclink{ACL2____MACROS}{macros} that allow us to quickly define
collections of related functions and theorems.  Some of these macros are widely
applicable, e.g., the \Xlibname{std/util} library's \texttt{defaggregate} macro
introduces something like a \texttt{struct} in C.  Other macros are very
customized and tailored to specific uses, e.g., the \Xlibname{VL} library's
\texttt{def-vl-modgen} macro is used to define a function that dynamically
creates a Verilog module and its supporting modules, according to a certain
paradigm.  A simple search for ``\texttt{(defmacro def}'' yields over 500
matches in the Community Books.  While exploiting guards and heuristically
identifying return-value theorems might allow us to document the \emph{output}
of these macros automatically, the abstraction itself would be lost.

Fortunately, there are good solutions to these problems.  In the rest of this
section, we'll explore the tools that XDOC provides for structuring books and
how macros can be extended to generate high-quality documentation
automatically.

\subsection{Automating the Topic Hierarchy}

Even without some kind of module system, it is still very common for closely
related definitions to be put together within the same book.  In a
well-structured project, the way that events are organized into books might
often be a good way to organize them in the manual.  This can be tedious to
implement.  The documentation hierarchy is built from the parents of
each individual topic, so to accomplish this organization, we may need to set
the \Xkwd{:parents} of each topic in our book to some common parent.

To make this kind of organization simpler to implement, the \Xdef{defxdoc}
command allows the \Xkwd{:parents} of a topic to be specified implicitly, using
the \Xtopiclink{ACL2____SET-DEFAULT-PARENTS}{xdoc::set-default-parents}
command.  For instance, the \Xtextlink{ACL2____OSLIB}{\Xlibname{oslib}} library provides various functions
for interacting with the operating system.  The logical definitions and
documentation for most of these functions are located in the same file.
Instead of specifying the parents explicitly for each topic, we can simply set
a default for the whole file.  For instance:




\begin{lstlisting}[language=Lisp]
(^\Xdef{in-package}^ "OSLIB")
(include-book ^\Ycmt{\dots}^)
(^\Xkwd{local}^ (xdoc::set-default-parents ^\Xfn{oslib}^)) ^\Ycmt{;; all topics below use :parents (oslib)}^

(^\Xdef{defxdoc}^ ^\Xfn{getpid}^
  ^\Xkwd{:short}^ "Get the current process identification (PID) number."
  ^\Xkwd{:long}^ ^\Ycmt{\dots}^)
(^\Xdef{defun}^ ^\Xfn{getpid}^ ^\Ycmt{\dots}^)

(^\Xdef{defxdoc}^ ^\Xfn{ls-subdirs}^
  ^\Xkwd{:short}^ "Get a subdirectory listing."
  ^\Xkwd{:long}^ ^\Ycmt{\dots}^)
(^\Xdef{defun}^ ^\Xfn{ls-subdirs}^ ^\Ycmt{\dots}^)
\end{lstlisting}

\subsection{Automating Event Listings}

Another very useful macro that comes with XDOC is named \Xdef{defsection}.
Aside from documentation, \Xdef{defsection} is similar to \texttt{(encapsulate
  nil ...)}; it provides a new scope for \Xtopiclink{ACL2____LOCAL}{local}
events, and usually will include at least some non-local events.  Like a
\texttt{defxdoc} command, it creates a new topic and lets you specify the
\Xkwd{:parents} and provide \Xkwd{:short} and \Xkwd{:long} strings for the new
topic.

But beyond this, \Xdef{defsection} can also automatically collect up the
non-local events within it, and inject them into the resulting documentation
topic.  Here is a basic example, taken from the classic \Xlibname{arithmetic}
library:

\begin{minipage}[t]{3.0in}
\vspace{0pt}
\begin{lstlisting}[language=Lisp]
(^\Xdef{defsection}^ ^\Xfn{inequalities-of-sums}^
  ^\Xkwd{:parents}^ (arithmetic-1)
  ^\Xkwd{:short}^ "Basic normalization to move
 negative terms to the other side of an
 inequality."

  (^\Xdef{defthm}^ ^\Xfn{<-0-minus}^
    (^\texttt{equal}^ (< 0 (- x))
     ^\texttt{~~~~~}^      (< x 0)))

  (^\Xdef{defthm}^ ^\Xfn{<-minus-zero}^
    (^\texttt{equal}^ (< (- x) 0)
     ^\texttt{~~~~~}^    (< 0 x)))

  (^\Xdef{defthm}^ ^\Xfn{<-0-+-negative-1}^ ^\Ycmt{\dots}^)
  (^\Xdef{defthm}^ ^\Xfn{<-0-+-negative-2}^ ^\Ycmt{\dots}^)
  (^\Xdef{defthm}^ ^\Xfn{<-+-negative-0-1}^ ^\Ycmt{\dots}^)
  (^\Xdef{defthm}^ ^\Xfn{<-+-negative-0-2}^ ^\Ycmt{\dots}^))
\end{lstlisting}
\end{minipage}
\hfill
\begin{minipage}[t]{3.5in}
\vspace{6pt}
~~\includegraphics[width=2.7in]{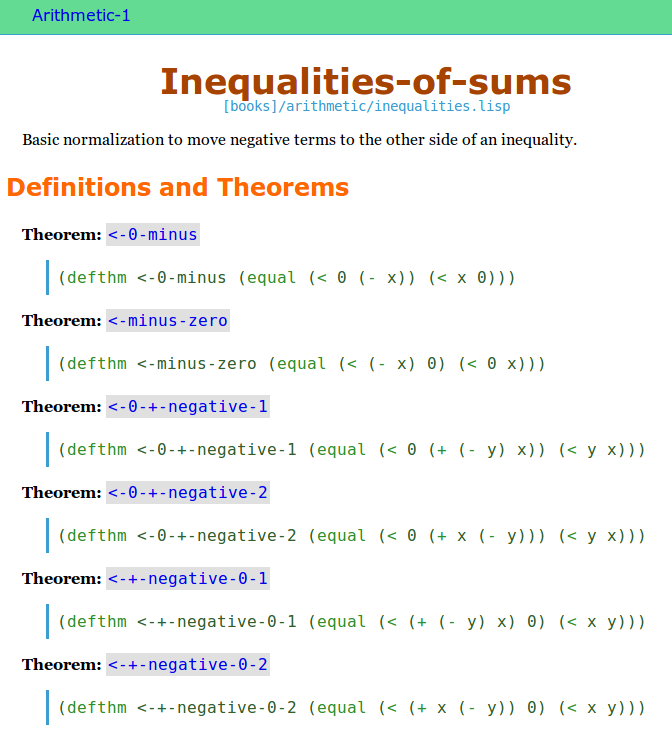}
\end{minipage}
\bigskip

The basic way that \Xdef{defsection} works is to collect up the names of the
events that have been submitted within it, and then convert them into
preprocessor directives.  For instance, for this example, it would create the
list \Xpp{@(def <-0-minus)}, \Xpp{@(def <-minus-zero)}, and so on.  These
directives are then added to the \Xkwd{:long} documentation provided by the
user, if any.  While we could do the same thing manually, this automation means
that the manual will automatically stay up to date when we add, remove, modify,
or reorder these theorems.




\subsection{Documentation-Aware Macros}

An important activity in large-scale ACL2 projects is the development of macros
to automate common patterns in modeling and proofs.  Each use of such a macro
might cause several interrelated functions to be defined, various tables to be
extended, and many theorems to be proved.

How can we effectively document the functionality created by these macros?  A
simple and effective answer is: extend the macro itself so that it also generates
appropriate \Xdef{defxdoc} forms.  Minimally, the macro might be augmented with
\Xkwd{:parents}, \Xkwd{:short} and \Xkwd{:long} arguments that can be used as
the basis for a top-level documentation topic.  More sophisticated and automatic
documentation may be possible if the macro also gives the user a way to add
documentation fragments in appropriate places.

As one example, \Xtextlink{STD____DEFAGGREGATE}{\Xdef{defaggregate}} is a macro in the \Xlibname{std/util} library
that can be used to introduce something like a \texttt{struct} in C.  It allows
documentation to be associated with each field.  Here is an example of using
this macro, taken from the \Xlibname{VL} library for working with Verilog:

\begin{lstlisting}[language=Lisp]
(^\Xdef{defaggregate}^ ^\Xfn{vl-assignstmt}^
  ^\Xkwd{:parents}^ (vl-stmt-p)
  ^\Xkwd{:short}^ "Representation of an assignment statement."
  ^\Xkwd{:tag :vl-assignstmt}^
  ^\Xkwd{:legiblep}^ nil

  ((type^~~^vl-assign-type-p
           "Kind of assignment statement, e.g., blocking, nonblocking, etc.")

   (lvalue vl-expr-p
           "Location being assigned to.  Note that the specification places
            various restrictions on lvalues, e.g., for a procedural assignment
            the lvalue may contain only plain variables, and bit-selects,
            part-selects, memory words, and nested concatenations of these
            things.  We do not enforce these restrictions in ^\Xpp{@('vl-assignstmt-p')}^,
            but only require that the lvalue is an expression.")

   (expr ^~^vl-expr-p
           "The right-hand side expression that should be assigned to the
            lvalue.")

   ^\Ycmt{\dots other fields \dots}^

   (loc  ^~~^vl-location-p
           "Where the statement was found in the source code."))

  ^\Xkwd{:long}^ "^\Xtag{<p>}^Assignment statements are covered in Section 9.2 of the Verilog
standard.  There are two major types of assignment statements:^\Xtag{</p>}^ ...")
\end{lstlisting}



\begin{figure}[tp]
\noindent 
{\tabulinesep=2.0mm
\begin{tabu}{l|l}
\includegraphics[width=3in]{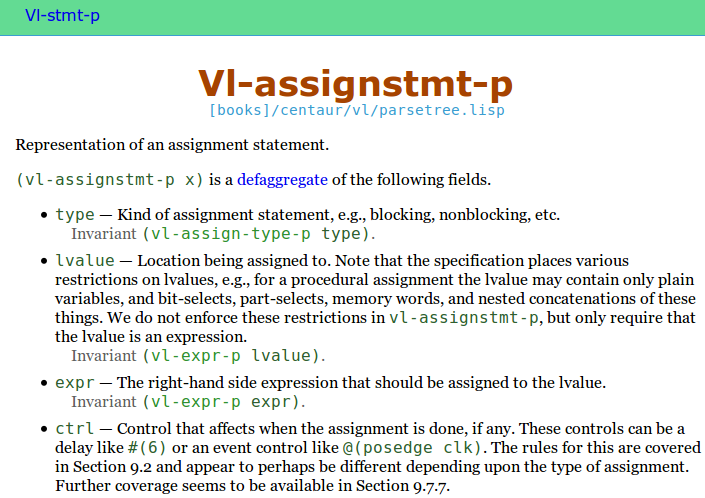}
&
\includegraphics[width=3in]{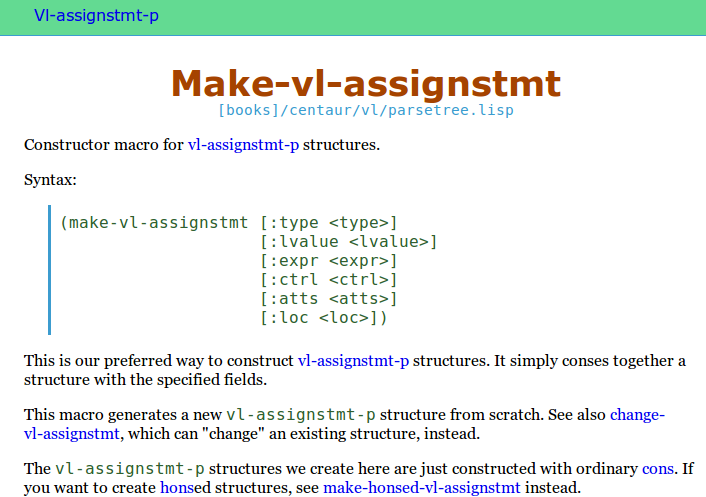}
\\
\hline
\includegraphics[width=3in]{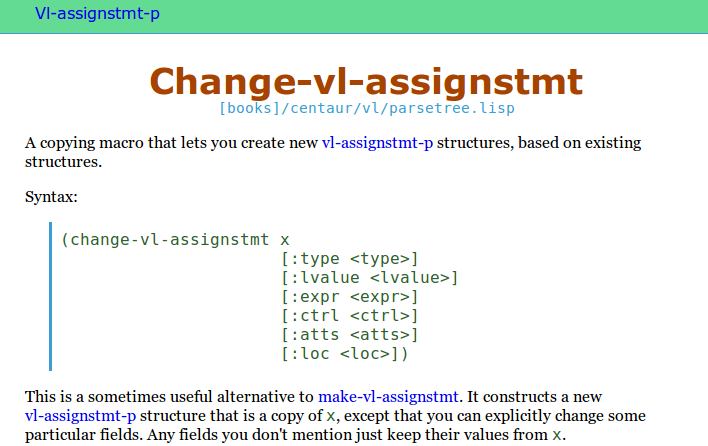}
&
\includegraphics[width=3in]{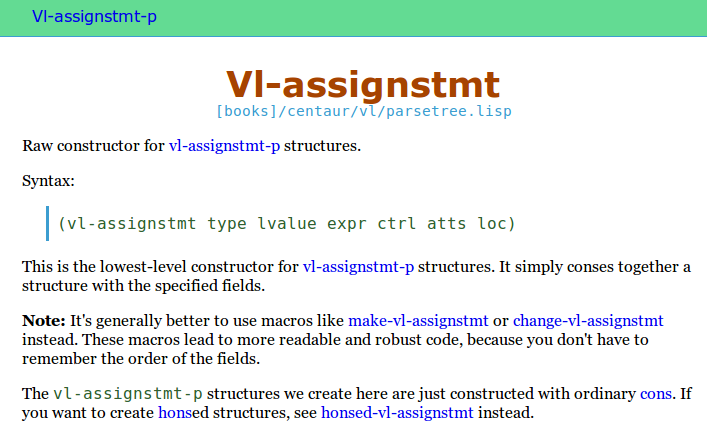}
\end{tabu}}

\caption{Some of the topics generated by \texttt{defaggregate} for \texttt{vl-assignstmt} structures.}
\label{fig:defaggregate}
\end{figure}

This use of \Xdef{defaggregate} produces several ACL2 functions and macros.  It
creates a recognizer named
\Xtextlink{VL____VL-ASSIGNSTMT-P}{\Xfn{vl-assignstmt-p}}, a constructor called
\Xtextlink{VL____MAKE-VL-ASSIGNSTMT}{\Xfn{make-vl-assignstmt}}, accessor
functions with names like
\Xtextlink{VL____VL-ASSIGNSTMT-_E3TYPE}{\Xfn{vl-assignstmt->type}}, and so on.
Each of these automatically becomes a corresponding topic in the manual; some
examples are shown in Figure \ref{fig:defaggregate}.  These new topics are
automatically grouped together with the recognizer being the main topic.  Even
if we don't add \emph{any} explicit documentation strings at all, our manuals
can at least include boilerplate documentation that, e.g., explains that this
structure is produced by \Xdef{defaggregate}, shows the fields of the structure
and the invariants they must satisfy, gives the syntax for the resulting
macros, and so on.  When we do provide additional documentation to describe the
fields of the structure, these explanations are nicely incorporated into the
manual.


A really elegant macro is \Xtopiclink{ACL2____DEFINE}{\Xdef{define}}, also from
the \Xlibname{std/util}
%
%
library.  This macro can be used to introduce, document, and prove theorems
about a new function.  Here is an example from the
\Xtextlink{ACL2____BITOPS}{\Xlibname{bitops}} library, which we will explain in
a moment:

\begin{lstlisting}[language=Lisp]
(^\Xdef{define}^ ^\Xfn{rotate-left}^
  ((^\texttt{x~~~~~}^ ^\texttt{integerp}^ "The bit vector to be rotated left.")
   (^\texttt{width~}^ ^\texttt{posp~~~~}^ "The width of ^\Xpp{@('x')}^.")
   (^\texttt{places}^ ^\texttt{natp~~~~}^ "The number of places to rotate left."))
  ^\Xkwd{:returns}^ (rotated natp ^\Xkwd{:rule-classes}^ ^\Xkwd{:type-prescription}^)
  ^\Xkwd{:short}^ "Rotate a bit-vector some arbitrary number of places to the left."
  ^\Xkwd{:long}^ "^\Xtag{<p>}^Note that ^\Xpp{@('places')}^ can be larger than ^\Xpp{@('width')}^....^\Xtag{</p>}^"
  (^\texttt{let*}^ ((width ^\texttt{~}^  (lnfix width))
   ^\texttt{~~~~~}^ (places ^\texttt{}^  (lnfix places))
   ^\texttt{~~~~~}^ (wmask  ^\texttt{~}^ (1- (ash 1 width)))
   ^\texttt{~~~~~}^      ^\Ycmt{\dots more bindings \dots}^
   ^\texttt{~~~~~}^ (ans    ^\texttt{~~~}^  (logior xl-shift xh-shift)))
   ^\texttt{~~}^ ans)
  ///    ^\Ycmt{ $\leftarrow$ special syntax: end of definition, start of related events}^
  (^\Xdef{defcong}^ ^\texttt{int-equiv}^ ^\texttt{equal}^ (rotate-left x width places) 1)
  (^\Xkwd{local}^ (^\Xdef{defthm}^ ^\Xfn{logbitp-of-rotate-left-1}^ ^\Ycmt{\dots}^)
  ^\Ycmt{\dots more theorems \dots}^
  (^\Xdef{defthm}^ ^\Xfn{logbitp-of-rotate-left-split}^ ^\Ycmt{\dots}^)
  (^\Xdef{defthm}^ ^\Xfn{rotate-left-by-zero}^ ^\Ycmt{\dots}^)
  ^\Ycmt{\dots more theorems \dots}^)
\end{lstlisting}

\smallskip

\begin{wrapfigure}[25]{r}[0pt]{3.2in}
\vspace{-1em}
\includegraphics[width=3.1in]{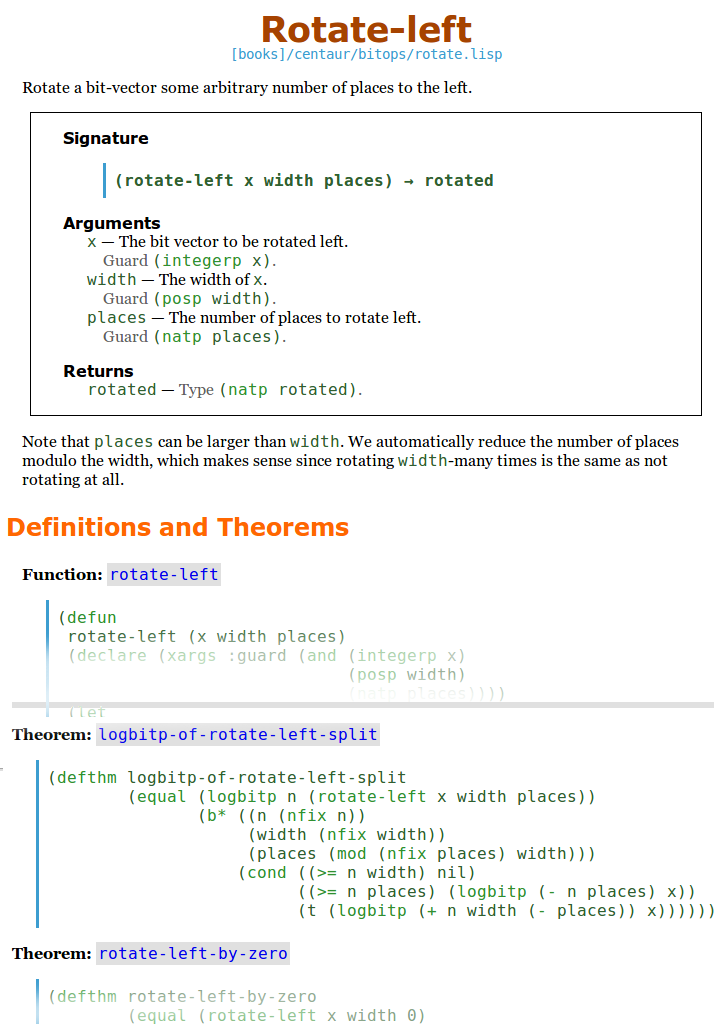}
\end{wrapfigure}

The \Xdef{define} macro connects to XDOC in three ways.  The most basic
is that it can be used as a replacement for
\Xtopiclink{COMMON-LISP____DEFUN}{defun} that simultaneously creates a new
topic, e.g., it provides the basic \Xkwd{:parents}, \Xkwd{:short}, and
\Xkwd{:long} arguments.

But when documenting a function, you very often want to describe its inputs and
return values.  To make this more convenient and automatic, \Xdef{define}
allows the arguments of the functions to be annotated with concise type
descriptions (which become guards) and
%
%
with documentation strings.  It also allows the return values of the function
to be named and documented, and provides a concise syntax for return-value
theorems.  This information is merged to create a
signature block in the manual that summarizes the function's interface, as
shown to the right.

Furthermore, when you introduce a function, it is very common to go on to prove
several theorems about it.  Using a special \texttt{///} notation,
\Xdef{define} provides a built-in area for related events about the new
function.  These events, along with the definition itself, are actually
submitted in a \Xdef{defsection} so that they can be automatically included in
the documentation.

Putting all of this together, \Xdef{define} does a great deal to minimize the
amount of redundant information between the code and documentation.
%
%
%
While coming up with good ways to integrate documentation strings into
macros can require some careful thought, it is not hard to write
macros like \Xdef{define}.  It is mostly a matter of collecting the
documentation from the various slots in the macro, and inserting it
into a template, e.g., using string concatenation.


\section{Conclusions}

We have developed XDOC, a mature and well-tested tool for documenting
formal verification efforts in ACL2.  We expect that every
organization that is using ACL2 for
large projects can benefit from developing an internal XDOC manual.

Note that this paper is not a comprehensive guide to using XDOC.
Practical details, such as the available tags in the
\Xtextlink{XDOC____MARKUP}{markup language}, the available
\Xtextlink{XDOC____PREPROCESSOR}{preprocessor directives}, and the
specifics of using various commands, can be found in XDOC's
documentation about itself; see the \Xtopiclink{ACL2____XDOC}{XDOC}
topic in the \XCombinedManual.


\subsection{Related Work}

XDOC treats ACL2 developments more like software than mathematics.  It has no
support for typesetting mathematical formulas or for explaining the details of
how proofs are carried out.  In contrast, a great deal of work has been done to
allow formal, mechanically checked proofs to be presented as if they were
traditional mathematics.  For instance,
\href{https://www.cl.cam.ac.uk/research/hvg/Isabelle/}{Isabelle} can produce
nicely typeset \LaTeX{} documents~\cite{haftmann-13-sugar} from its input files
(theories), and interfacing layers like
\href{http://isabelle.in.tum.de/Isar/}{Isar}~\cite{99-wenzel-isar} allow the
source code to be written in a style that is similar to paper-and-pencil
proofs.  Other efforts such as the
\href{http://www.ru.nl/foundations/research/projects/mathwiki/}{MathWiki}
project~\cite{11-alama-wikis} have also focused on develop Wiki systems for
formalized mathematics.
Garc\'{i}a-Dom\'{i}nguez~\cite{09-dominguez-hypertext} et al.~have also
developed tools for viewing and understanding ACL2 proofs.

XDOC's basic approach---embedding documentation fragments in the
code---is used in many popular software documentation generation tools
such as \href{http://www.oracle.com/technetwork/java/javase/documentation/javadoc-137458.html}{Javadoc},
\href{http://www.doxygen.org/}{Doxygen}, and
\href{http://qt-project.org/wiki/Category:Tools::QDoc}{QDoc}.  These
tools have long been used to provide high-quality manuals for large
software projects such as the
\href{http://docs.oracle.com/javase/7/docs/api/}{Java platform},
\href{http://qt-project.org/}{QT}, \href{http://api.kde.org/}{KDE},
and many others.  Similar tools like
\href{http://perldoc.perl.org/perlpod.html}{POD} for Perl,
\href{http://rdoc.sourceforge.net/}{RDOC} for Ruby, and so on, differ
in matters of markup and other details, but are also typically based
on documentation comments.  The \href{http://coq.inria.fr/}{Coq}
theorem prover has a similar tool,
\href{http://coq.inria.fr/refman/Reference-Manual017.html#coqdoc}{coqdoc},
that can produce \LaTeX{} and HTML.



Flatt, Barzilay, and Findler~\cite{09-flatt-scribble} have developed an
impressive documentation generation system for
\href{http://racket-lang.org}{Racket}, called
\href{http://docs.racket-lang.org/scribble/}{Scribble}, which is very different
from these comment-based tools.  It features a very interesting \texttt{@}
syntax that is closely related to S-expressions but is designed primarily for
writing text.  This syntax allows for quick transitions between S-expressions
and text in documentation.  Scribble allows for the documentation to
be manipulated by programs, and for the development of custom macros for use in
generating documentation.

Knuth has advanced Literate Programming~\cite{84-knuth-literate} as an
alternative to documentation generation systems, where instead of
embedding documentation in the code, code is embedded in the
documentation.  This approach allows the code itself to be structured
more like the manual, e.g., definitions might appear in a top-down
fashion in the source code file.  Gamboa~\cite{03-gamboa-literate} has
described an XML-based Literate Programming tool for developing and
presenting ACL2 proofs.  In this tool, ACL2 source code is extracted
from an XML document, perhaps in an out-of-order fashion, so that the
document can be arranged for human comprehension, e.g., in a top-down
fashion.  Separately, the document can be converted into other
formats, e.g., web pages.



\subsection{Future Directions}

Although the web-based XDOC viewer has many features, a few known bugs and
quirks have been reported in the
\href{https://code.google.com/p/acl2-books/issues}{issue tracker} for the
Community Books.  Its main missing feature is a good searching operation.  The
web-based viewer can currently search topic names and short strings.  Some care
has been taken to make this work well, and to feature ``important'' topics more
prominently in the search results.  But it would certainly be much better to
include the long strings in the search!  This seems tricky to implement
consistently across both server-supported and local manuals.  In the interim,
the \Xacldoc Emacs browser can carry out basic full-text substring and
regular-expression searches.

Integrated development environments such as Eclipse and Visual Studio provide
extremely useful ``intelligent code completion'' features for languages like
Java.  When a programmer who is using this feature begins to type, say, a
particular function call, he is automatically shown an unobtrusive reminder of
the function's signature and documentation.  It would be very exciting to have
this capability for ACL2 functions.  Perhaps XDOC, with its simple and widely
supported markup language, could serve as the data source for this kind of
functionality, e.g., in the ACL2 Sedan.

\subsection{Acknowledgments}

Any manual can only be as good as its content.  We have greatly benefited from
the many other people who have made substantial contributions to the
documentation of ACL2 and the Community Books.  Before 2014 (through ACL2
Version 6.4), the authors of ACL2 were responsible for all documentation in the
ACL2 User's Manual.  They wrote all of it except
for certain topics that are typically labeled with their author, and except for
the \Xtextlink{COMMON-LISP____REAL}{ACL2(r)},
 \Xtextlink{ACL2____HONS-AND-MEMOIZATION}{ACL2(h)}, and
\Xtextlink{ACL2____PARALLELISM}{ACL2(p)} documentation, which
includes contributions from Bob Boyer, Jared Davis, Ruben Gamboa, Warren Hunt,
David Rager, and Sol Swords.  ACL2's documentation was initially marked up by
Laura Lawless.  The documentation for the Community Books includes significant
contributions from so many authors that we will not try to list them here.



Many other people have made important contributions to XDOC itself.  Davis'
colleagues at Centaur, including Alan Dunn, Harsh Raju Chamarthi, Sol Swords,
and Anna Slobodov\'{a}, provided extensive beta testing, wrote significant
documentation, and gave us helpful feedback and ideas that became important
features.  Kevin Kramer and Patrick Roberts provided technical support related
to hosting XDOC manuals.  David Rager was an early adopter and
supporter of XDOC, and aside from his many useful suggestions, he has made
numerous corrections to many topics.

Warren Hunt was instrumental in pushing us to unify the legacy documentation
and XDOC systems.  Our conversion of the legacy documentation into XDOC format
was made much easier by the legacy documentation system's tools for creating
HTML and Emacs Info manuals.  These converters were initially developed by Art
Flatau and were subsequently improved by Michael Bogomolny and Alan Dunn.  J
Moore provided useful insights during this conversion process, and also
provided extensive beta-testing.

We thank David Rager, Sol Swords, and the workshop reviewers for
feedback on drafts of this paper.

This material is based upon work supported by DARPA under Contract
No. N66001-10-2-4087, by ForrestHunt, Inc., and by the National
Science Foundation under Grant No. CCF-1153558.

\bibliographystyle{eptcs}
\bibliography{paper}
\end{document}